\documentclass[trackchanges, twocolumn]{aastex701}
\usepackage{amsmath}

\newcommand\chas{CH$\alpha$S}
\newcommand\ha{H$\alpha$}
\newcommand\hb{H$\beta$}

\begin{document}

\title{Identifying signatures of inflow onto face-on galaxies using the Balmer decrement}

\author{Meghna Sitaram}
\affiliation{Department of Astronomy, Columbia University, New York, NY 10027, USA}
\email[show]{m.sitaram@columbia.edu}  

\author{Hui Li} 
\affiliation{Department of Astronomy, Tsinghua University, Beijing 100084, People's Republic of China}
\email[show]{hliastro@tsinghua.edu.cn}

\author{Yong Zheng}
\affiliation{Department of Physics, Applied Physics and Astronomy, Rensselaer Polytechnic Institute, Troy, NY 12180, USA}
\email{zhengy14@rpi.edu}

\author{Greg L. Bryan}
\affiliation{Department of Astronomy, Columbia University, New York, NY 10027, USA}
\email{gb2141@columbia.edu}

\author{Mary Putman}
\affiliation{Department of Astronomy, Columbia University, New York, NY 10027, USA}
\email{mep2157@columbia.edu}

\author{Aaron Smith}
\affiliation{Department of Physics, The University of Texas at Dallas, Richardson, Texas 75080, USA}
\email{asmith@utdallas.edu}

\author{Rahul Kannan}
\affiliation{Department of Physics and Astronomy, York University, 4700 Keele Street, Toronto, ON M3J 1P3, Canada}
\email{kannanr@yorku.ca}

\begin{abstract}

Isolated star-forming galaxies require inflows of fresh gas from the surrounding medium to sustain episodes of star formation over time. 
However, there are very few direct detections of accretion onto external galaxies. Studies in absorption can only observe along limited sightlines, while those in emission can have difficulty distinguishing inflowing gas in the foreground of the galactic disk from similarly Doppler-shifted outflowing gas in the background. We explore the possibility of using the Balmer decrement (H$\alpha$/H$\beta$) in low-inclination systems as a diagnostic for disentangling the flow geometry in disk-like galaxies. We leverage mock spatial-spectral observations of an isolated Milky Way-mass galaxy simulated using the radiation-hydrodynamics code AREPO-RT and post-processed with the Monte Carlo radiative transfer code COLT. We find that gas components located in front of the disk exhibit systematically lower Balmer decrements than gas embedded in or behind the disk, with a mean front-back offset of $\Delta(\text{H}\alpha/\text{H}\beta) \approx -0.14$. The ability to differentiate between the disk and far-side components is limited by the extremely clumpy, multiphase dust distribution along the line of sight introducing substantial scatter. Overall, the results provide a useful observational diagnostic of inflow and outflow in dusty face-on galaxies.

\end{abstract}

\keywords{\uat{Galaxies}{573} --- \uat{Galaxy kinematics}{602} --- \uat{Galaxy accretion}{575} --- \uat{Galaxy Spectroscopy}{2171}}


\section{Introduction} 

The cycling of gas between galaxies and their surrounding media is an important aspect of the formation and evolution of galaxies. Gas inflow in particular is a crucial yet elusive process with very few direct detections to date. There is plentiful indirect evidence supporting the presence of gas inflow and accretion throughout cosmic time \citep{Putman2017}.  
In the Milky Way, the star formation rate has been constant, estimated at around 1.5 to 2.5 M$_{\odot}$ yr$^{-1}$ \citep{chomiuk2011, licquia2015, elia2022} for the past several Gyr \citep{chiappini1997}; however, the gas mass of the Galaxy indicates that this rate could only be supported for a few Gyr. For galaxies at a variety of redshifts, star formation rates suggest that galaxies will deplete their gas supply in times ranging from only 0.7--2.2 Gyr without an ongoing influx of gas from the surrounding circumgalactic medium (CGM) and intergalactic medium (IGM) \citep{schiminovich2010, leroy2013, genzel2010, tacconi2013}. In the Milky Way, observational evidence for inflow includes the detection of high- and intermediate-velocity clouds infalling onto the disk \citep{lehner2010,Clark2022}. Also, the metallicity distribution of G stars through the galaxy is consistent with models including replenishment of star-forming material from pristine inflows \citep{chiappini2009, fenner2003}.  

Despite the indirect evidence suggesting the need for gas
inflow, and the observational evidence in the Milky Way, there are only a few direct detections of inflowing gas in other galaxies. 
One of the strongest detections comes from \cite{zheng2017}. The authors used the Cosmic Origins Spectrograph on the Hubble Space Telescope to obtain UV spectra of 7 bright stars in M33, a nearby galaxy with an inclination of 56$^{\circ}$. At this inclination, inflowing gas will have significant velocity along the line of sight of observations. Investigation into multiple UV absorption lines 
revealed a redshifted component separated  
from the rest velocity of the galaxy by about 110\,km\,s$^{-1}$. This study used UV-bright stars within the galaxy as background sources for absorption, ensuring that any absorbers would be in front of the galaxy, and red-shifted components would then be inflowing onto the disk. At higher redshifts, additional cases of inflow have been detected at large scales using the galaxy itself as the background source \citep{rubin2012}, on smaller scales for inflowing absorbers that fortuitously fall in front of a bright source in the disk \citep{weldon2023, coleman2024}, or by using kinematic modeling to determine whether gas is inflowing or outflowing \citep{bouche2016}. 
Absorption measurements allow for high signal-to-noise detections of faint gas (such as \cite{james2022}), but they are limited to a few sightlines with bright background sources or UV-bright stars, and therefore cannot explore the spatial distribution of inflows and outflows. Observations using gravitational arc tomography can measure the density of gas in absorption over a wider spatial extent of the CGM halo, but depend on kinematic modeling to determine the position of gas relative to the galaxy \citep{lopez2020, tejos2021, mortensen2021}.
Observations in emission can provide measurements of gas flows over an entire galaxy. At low-redshifts, strong UV lines commonly used to probe ionized gas are inaccessible from the ground, and optical transitions such as the optical Balmer emission lines become faint but valuable tracers \citep{corlies2016, lokhorst2019}.

When observing in emission, each spectrum  encodes the emission intensity and the Doppler velocity of the gas. This can create confusion as to the location of different kinematic components along a sightline.  
For example, in a low-inclination galaxy, inflowing gas in front of the galaxy and outflowing gas observed through the galaxy would both appear as a spectral component redshifted compared to the disk component. From the spectrum alone, it is difficult to disentangle where this component falls with respect to the disk on the sightline. Previous IFU studies which have observed accreting gas in emission rely on additional absorption observations or models of galaxy kinematics to judge whether gas is inflowing or outflowing \citep{johnson2022, vayner2023, bolda2025}.
Other observational studies have assumed that all gas at velocities offset from the ISM represent outflows, or that all offset spectral components represent gas from the front of a galaxy \citep{mcquinn2019, reichardtchu2025}. However, spectrum wings have displayed different reddening values with respect to the ISM depending on the galaxy sample \citep{mingozzi2019, Fluetsch2021, villarmartin2014, baron2024}. 
In this work, we investigate the possibility of using the Balmer decrement (the ratio between \ha{} and \hb{} emission) and its dependence on extinction, to break this degeneracy and identify inflow and outflow in \ha{} and \hb{} emission spectra, without the need for additional observations. 

\begin{figure*}
    \includegraphics[width=\textwidth]{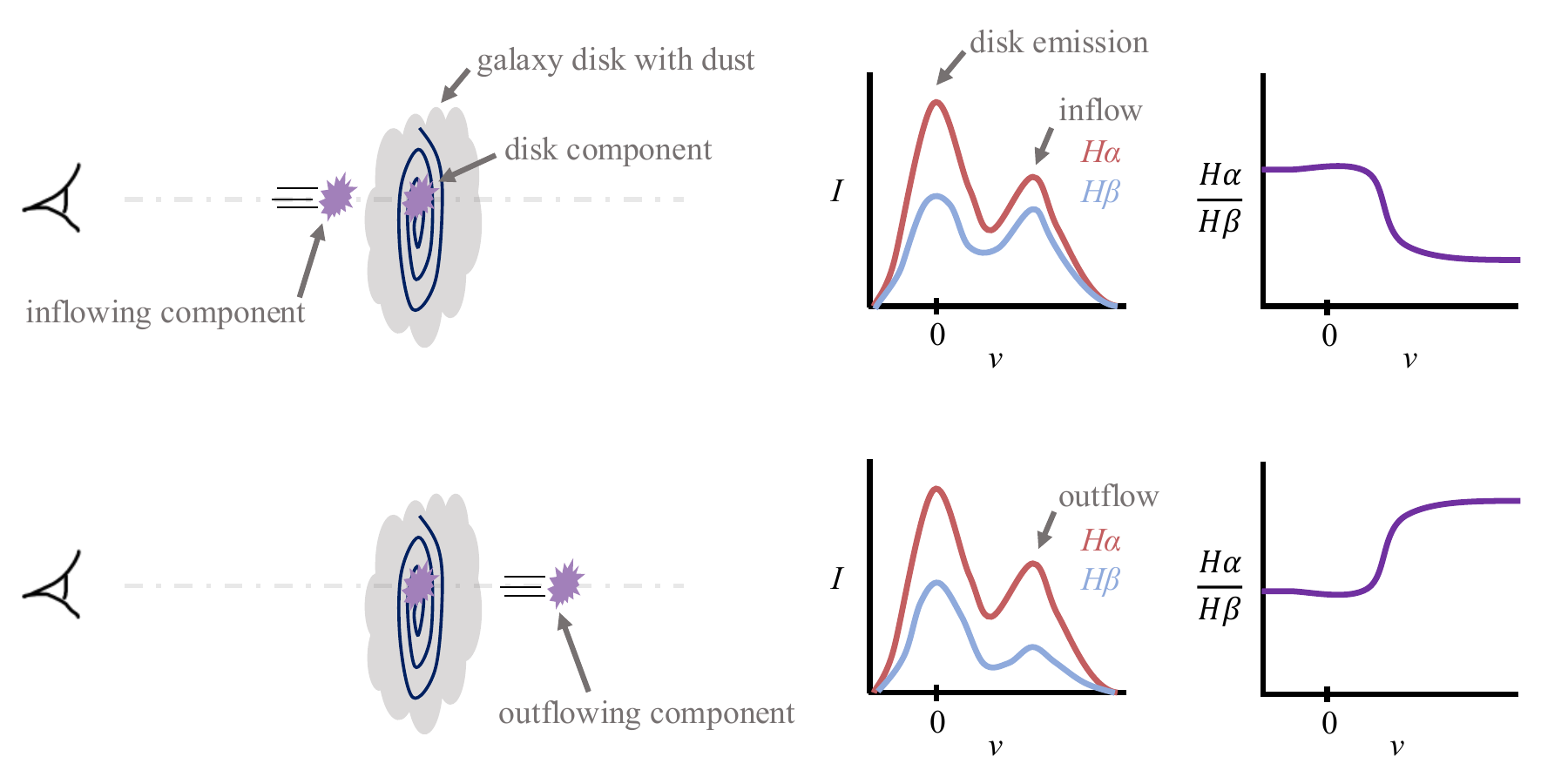}
    \caption{An exaggerated schematic displaying the degeneracy in emission spectra of galaxies. In the top panel, an inflowing component in front of the galaxy appears as a component redshifted from the disk component in the spectrum. In the bottom panel, an outflowing component behind the galaxy similarly appears as a redshifted component. Though the \ha{} emission is similar, the \hb{} emission will be more affected by extinction due to dust in the galaxy. Therefore, the Balmer decrement will reveal which component is inflowing and which is outflowing. }
    \label{fig:cartoon}
\end{figure*}

The Balmer decrement has historically been used to estimate the total attenuation due to dust in galaxies. Electrons in a hydrogen atom have intrinsic transition rates between energy levels which partly determine the ratio between \ha{} and \hb{} emission. 
For gas with temperatures between $5\times10^{3}$ and $10^{4}$\,K, roughly the temperature of \ha{} emitting warm gas, the expected \ha{} to \hb{} emission ratio will be 2.86--3.04 \citep{groves2011}. However, dust lying between the emitting gas and the observer will absorb some emitted photons, and will preferentially absorb bluer wavelengths, raising the \ha{}/\hb{} ratio.
Variation in the Balmer decrement in the same spectrum for different kinematic components may help locate these components along the sightline and in relation to the galaxy. This is illustrated in Figure \ref{fig:cartoon}. In the top panel, there is a component in front of the galaxy moving away from the observer and into the galaxy. In the bottom panel, the component is observed through the galaxy moving away from both the observer and the galaxy. The middle panels display how these cases yield very similar \ha{} spectra. However, as seen in the right panels, gas from in front of the galaxy will have traveled through very little dust compared to the ISM, resulting in a lower \ha{}/\hb{} ratio. Meanwhile, gas from behind the galaxy will have traveled through more dust on its path to the observer, decreasing the \hb{} emission and increasing the \ha{}/\hb{} ratio of that component.

To explore and test this method, we use mock observations of an AREPO-RT simulation of an isolated Milky Way-like galaxy with state-of-the-art multiphase ISM and stellar feedback physics modeling \citep{kannan2020}. As a middle ground between cosmological-size and small-scale simulations, this simulation contains an entire galaxy with inflowing and outflowing gas,  
detailed star formation, and dust effects. Importantly, extinction is included, allowing detailed studies of the ionized gas kinematics, and the resulting spectra  
with the Balmer decrement. 
We use these mock observations to investigate whether and how the Balmer decrement can be used to identify the $z$-positions of spectral components along a sightline. 
In Section \ref{sec:simulation}, we introduce the simulation and radiative transfer and describe the characteristics of the simulated galaxy. In Section \ref{sec:methods}, we recount the methods used to determine the positions of different spectral components along the sightline. In Section \ref{sec:results}, we start by describing a case study of a single sightline exhibiting inflow, and then discuss the results for all sightlines. In Section \ref{sec:discussion}, we discuss how the dust distribution and different aspects of the simulation affect this method and our result. Finally, in Section \ref{sec:summary}, we summarize the paper.

\section{Simulation} \label{sec:simulation}

\begin{figure*}
    \includegraphics[width=\textwidth]{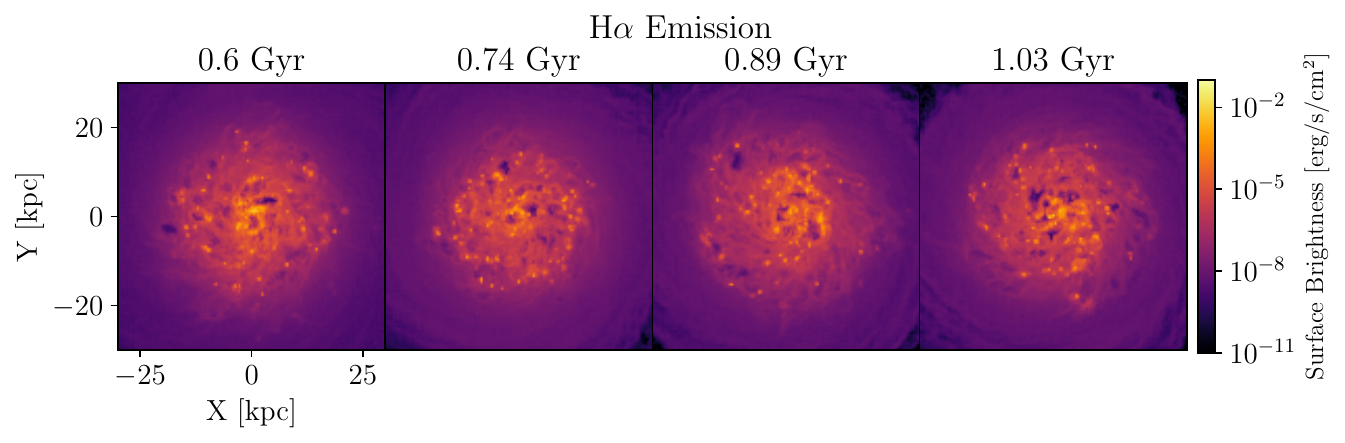}
    \caption{\ha{} emission maps of the simulation for four different snapshots equally spaced in time after reaching quasi-steady state, output by the radiative transfer code COLT. These four simulation snapshots are used for the rest of the analysis.  We note the similar overall structure between the four time-steps, allowing us to apply a similar analysis to each timestep.}
    \label{fig:haemission}
\end{figure*}

To search for signatures of inflow and outflow in \ha{} and \hb{} emission lines, we analyze a high-resolution radiation hydrodynamical simulation of an isolated Milky Way-sized galaxy \citep{kannan2020}. This simulation was performed using AREPO-RT \citep{kannan2019}, a radiative transfer extension on the moving-mesh hydrodynamic code AREPO. We refer the reader to  \citet{marinacci2019} and \citet{kannan2020} for a description of the adopted sub-grid models for star formation and feedback.  The simulation includes many key physical ingredients, such as non-equilibrium thermochemistry, dust formation and destruction, star formation, stellar feedback from radiation, stellar winds, and SNe implemented as self-consistently as possible given the target resolution and galaxy mass. Dust is evolved self-consistently following the empirical formation–destruction framework of \citet{McKinnon2017}.  In particular, dust is treated as a passively advected scalar and its mass is tracked separately for five chemical species, assuming a fixed grain size of $a=0.1\,\mu\mathrm{m}$. Production occurs through mass return from SN\,II, SN\,Ia, and AGB stars, while interstellar grain growth proceeds via gas-phase metal accretion on a characteristic timescale $\tau_{\rm g}\propto \rho^{-1}T$. Destruction is governed by a combination of supernova shock processing, parameterized by a destruction timescale set by the local SN\,II rate, and thermal sputtering at high temperatures. This methodology yields spatially variable dust-to-gas ratios that respond self-consistently to local density, temperature, and radiation fields. The simulation therefore exhibits a realistic multi-phase interstellar medium (ISM) and large-scale galactic winds triggered by stellar feedback. 

This is a non-cosmological simulation, modeled after low-redshift systems, with initial conditions containing a dark matter halo ($1.5\times10^{12}M_\odot$), a stellar disk+bulge ($6.2\times10^{10}M_\odot$), and gaseous disk ($4.2\times10^{9}M_\odot$), consistent with that of the Milky Way. The simulation is run with a stellar mass resolution of $2.8 \times 10^3$ M$_\odot$ and a gas mass resolution of $1.4 \times 10^3$ M$_\odot$. The corresponding gravitational softening length is $\epsilon _{star} = 7.1$ pc and the simulation is run for approximately 1 Gyr. The star formation rate (SFR) of the model galaxy is maintained around 2--4\,M$_\odot$\,yr$^{-1}$, consistent with the current status of our Galaxy \citep{chomiuk2011}. 

To generate synthetic \ha{} and \hb{} emissions, we perform post-processing Monte Carlo radiative transfer calculations on the simulation snapshots at different epochs employing the Cosmic Ly-$\alpha$ Transfer code \footnote{See \url{colt.readthedocs.io} for public access and documentation.}\citep[COLT;][]{Smith2015, smith2021}. We utilize the recalculated hydrogen ionic abundances in ionization equilibrium with the stellar radiation field, based on age--metallicity tabulated spectral energy distributions from the high-resolution \citet{BruzualCharlot2003}, allowing direct gas reprocessed Balmer line emission arising from recombination and collisional excitation mechanisms. COLT delivers datacubes of spatially- and spectrally-resolved maps, that is in different velocity or frequency channels for both \ha{} and \hb{}. It also allows us to explore different dust models, such as pure absorption or including dust scattering; in either case we adopt the fiducial Milky Way dust model from \citet{WeingartnerDraine2001}. The rich information in the mock dataset is thus a powerful tool to probe inflow and outflow at different locations in the simulated galaxy and provide guidance and interpretation to future observations. One output of the COLT simulations is displayed in Figure \ref{fig:haemission}, which shows observed \ha{} emission maps of the simulated galaxy face-on in four simulation snapshots which are used in the remaining analysis.

\subsection{Galaxy Characteristics} \label{subsec:galchar}

\begin{figure*}
	\includegraphics[width=\textwidth]{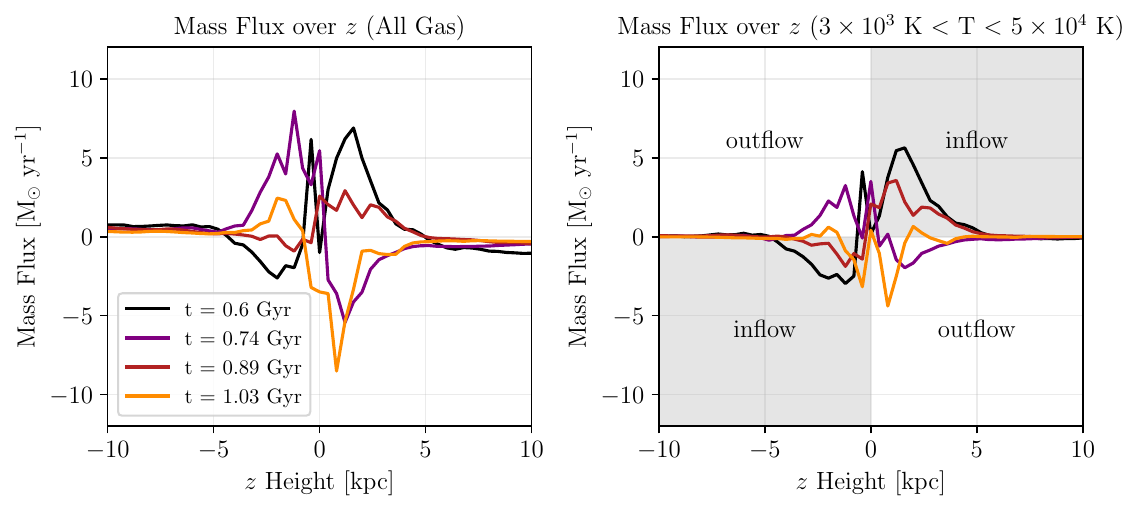}
    \caption{Left: The mass flux of different slices of the galaxy at different $z$ height above and below the disk. Four different snapshots of time are shown. Right: The same but only for \ha{}-emitting gas between $3 \times 10^3$\,K and $5 \times 10^4$\,K. Regions of the plot corresponding to inflow and outflow are displayed on the right panel.}
    \label{fig:massfluxgraph}
\end{figure*}

To ensure that the simulated galaxy hosts realistic dust and a variety of inflows and outflows, we examine the ionized gas mass flux and dust mass across the galaxy. Figure \ref{fig:massfluxgraph} displays the summed mass flux at different heights through the galaxy at each of the four timesteps. 
The left hand panel displays the mass flux for all gas temperatures across the galaxy at different heights above and below the disk. We calculate the total mass flux by summing the multiplied $z$-velocity and mass of each particle in a slice of 0.2\,kpc thickness at various heights above and below the disk. For particles which fall only partly inside the slice, only the mass included in the volume intersecting the slice is summed (assuming that the particle is spherical).  As seen in the left panel, bursty star formation events create large outflows in some timesteps. One example occurs at $t=0.74$ Gyr (and again at 1.03 Gyr), in which outflow can be seen on both sides of the disk. Between these periods (i.e., at $t=0.6$ Gyr and 0.89 Gyr), inflow can be seen.

The right panel displays the mass flux of the warm gas that will emit \ha{} and \hb{} for the same timesteps. Warm gas was identified as gas with temperatures between $3\times10^3$ K and $5\times10^4$ K. The warm gas largely follows the flows observed for all gas, including some large-scale flows due to energetic events like star formation. Each snapshot shows examples of inflows and outflows for which the Balmer decrement can be studied. This simulation is of an isolated Milky Way galaxy, without a surrounding intergalactic or circumgalactic medium. As such, the inflows and outflows identified represent a process more like recycling through the galaxy rather than inflow from the IGM. However, the method being tested for identifying inflows onto the galaxy should be similar regardless of the source of the inflow or outflow, although there would be differences in the mass of gas, metals, and dust in these flows.

\begin{figure*}
\centering
	\includegraphics[width=0.9\textwidth]{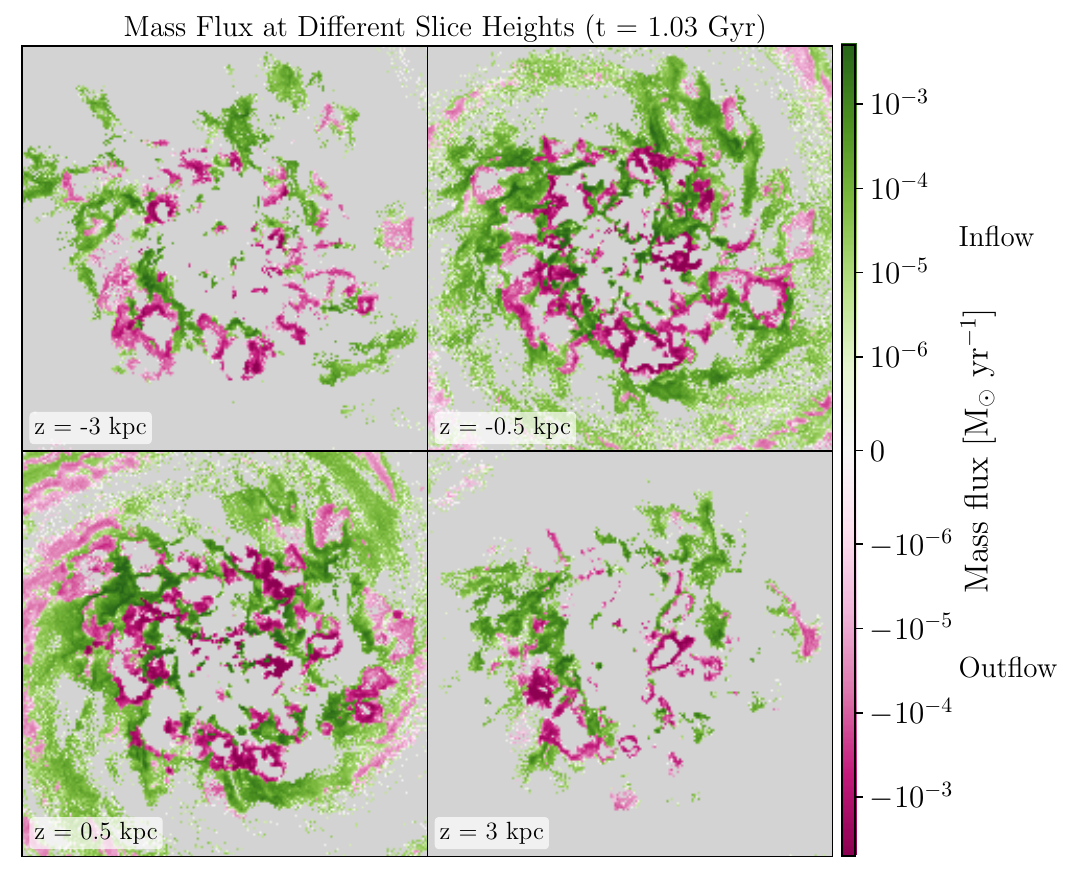}
    \caption{All panels show colormaps of the mass flux of gas at temperatures between $3 \times 10^3$ and $5 \times 10^4$ K across a slice of the galaxy with slice height labeled on the bottom right. Green corresponds to inflow and pink corresponds to outflow. Gray regions show pixels where there are no warm gas particles. }
    \label{fig:massfluxcolormap}
\end{figure*}

In Figure \ref{fig:massfluxcolormap}, we show the spatial distribution of the mass flux of warm gas across the galaxy at different $z$-heights between $-3$\,kpc and 3\,kpc. All slices are taken from the snapshot at timestep $t=1.03$ Gyr, which is the latest simulation snapshot and a time when the warm gas shows mixed inflow and outflow. In this figure, green areas represent inflowing gas while pink areas represent outflowing gas. Gray regions have no particles between $3\times10^3$ K and $5\times10^4$ K. 
There is significant small-scale structure visible in the mass flux, with inflow and outflow mixed in many regions. Even at heights of $z = 3$\,kpc, in which there is supposed to be strong outflow (as shown in Figure \ref{fig:massfluxgraph}) depicted in the dark pink regions, there are still significant regions of inflow mixed in. 
As we move away from the disk in height, the levels of inflow and outflow decrease as expected, as there is less warm gas at large heights above and below the disk than in the disk. 
The existence of mixed inflow and outflow at different $z$-heights across the galaxy allows us to study how the Balmer decrement can change even within the same sightline due to the locations of the emitting gas, and whether there are trends in this change that can be used to guide observations.

\section{Methods} \label{sec:methods}

In Section \ref{subsec:datacube}, we describe how mock spectra were created from the simulated galaxy and traced back to the simulation particles. Then, Section \ref{subsec:methods2} describes how we found the $z$-positions of the emitting particles responsible for different components of the spectra. 

\subsection{Data Cube Generation} \label{subsec:datacube}

As previously mentioned, COLT was used to self-consistently generate spectra of the \ha{} and \hb{} emission lines from the galaxy. The primary data products we work with are individual data cubes, for \ha{} and \hb{}, each with a spatial resolution of 30\,pc and a velocity resolution of 5\,km\,s$^{-1}$. The field of view covers out to 30\,kpc from the center of the galaxy and a range of $-400$ to 400\,km\,s$^{-1}$ in line-of-sight velocity. The data cube also includes different options for inclusion of physical effects of dust on the spectra. In our analysis, we make use of the \textbf{``intrinsic''} spectra which exclude scattering and absorption/reddening, the \textbf{``reddened''} spectra which include absorption but exclude scattering (i.e. zero scattering albedo), and the \textbf{``processed''} spectra which include both physical effects. All data are based on capturing the observed photon properties and next-event estimation (peel-off) calculations \citep{Yusef-Zadeh1984}, which improves Monte Carlo statistics by capturing the potential contributions of all scattering events from all photon packets after correcting for $e^{-\tau}$ attenuation to the observer.

Since the intrinsic spectra provided by COLT ignore the effects of dust, the intensities and frequencies of the spectra should depend only on the luminosities and velocities of the emitting particles. To cross-check the velocity cubes to the simulation, and ensure a correct match from the spectrum to the emitting particles, we tested whether we were able to recreate the intrinsic spectra using particle information only. For each sightline, we identified the particles inside using their position. We included only particles whose centers fell inside the sightline.
As a first check, we compared the sum of the particle luminosities (using the same calculations as COLT) to the sum of the intrinsic spectra. Next, we recreated the complete intrinsic spectrum by making a luminosity-weighted histogram with velocity bins of 5\,km\,s$^{-1}$, matching the spectral resolution of the COLT data cube. We smoothed the histogram with a Gaussian kernel of width 10\,km\,s$^{-1}$, which is within the expected range of Doppler widths of gas at \ha{}-emitting temperatures.

Our objective is to identify which peaks in the galaxy spectra correspond to inflows and outflows by determining the positions and velocities of the particles contributing to each spectral component in the data cube. This is complicated by two effects: 1) simulation particle sizes are often at the same scale as the data cube's native resolution of 0.03\,kpc, and 2) areas with very low flux appear noisier compared to other regions due to the Monte Carlo method used for the radiative transfer. 
Co-adding spectra over larger regions helps to reduce the noise in low-emitting regions and bring the spatial resolution closer to the same order as the usual resolution observable by spectrographs which search for CGM emission.  
It also increases the spectral resolution above the size of most particles, clarifying which particles contributed their luminosities to which sightline. 
Because of this limitation in particle identification, larger spatial resolution will yield more accurate intrinsic spectra. As binning increases and spatial resolution decreases, there will be many more particles whose entire volumes fall inside a single sightline, making the sum of their luminosities more accurate. The optimal spatial resolution for the remainder of the analysis is one that would be comparable to observations while still allowing traceability between a spectrum and the particles, and therefore any inflow and outflow. 

We tested spatial resolutions ranging from 0.03\,kpc (the pixel size of the data cube) to 1.5\,kpc ($50 \times 50$ data cube pixels).  Using this method, we were able to completely and accurately identify the particles at spatial resolutions of 0.3\,kpc ($10 \times 10$ binning). This spatial resolution is comparable to current IFU spectrographs.  The spatial resolution of the Circumgalactic \ha{} Spectrograph (\chas{}) is 1--5\,kpc for a galaxy at $d = 10$\,Mpc \citep{melso2022}. This is a new narrowband integral field unit spectrograph designed to target faint emission (so far, \ha{}, \hb{}, and O III) in the disk and CGM of nearby galaxies. 
The MaNGA galaxy survey also samples its target galaxies at spatial scales of approximately 1--2\,kpc \citep{drory2015}. Some IFUs do have finer spatial resolution, such as VIRUS and MUSE at $\approx 48$\,pc and 14--19\,pc respectively for galaxies at a distance of 10\,Mpc \citep{hill2004, bacon2010}.   
However, for our data set, as the spatial resolution decreases past 0.3\,kpc, more simulation particles become of a comparable size to the sightline itself, causing discrepancies between the intrinsic spectrum calculated using this particle-based method and the results of the radiative transfer.

\subsection{Tracking inflows and outflows} \label{subsec:methods2}

From the simulation data cubes, we build up a sample of sightlines containing inflow and outflow across the galaxy and measure their Balmer decrements. 
In order to further reduce noise in low \ha{}-emitting regions, we apply a spectral flux limit based on the observability limits of \chas{}. \chas{} can observe emission down to 2706\,photons\,s$^{-1}$\,cm$^{-2}$\,sr$^{-1}$ \citep{melso2022}. When converted based on the photon energy at \ha{}, the 0.3\,kpc $\times$ 0.3\,kpc size of the sightline, and the velocity bins of the data cube, the spectral flux limit is $2.06 \times 10^{-8}$\,erg\,s$^{-1}$\,cm$^{-2}$/(km\,s$^{-1}$). The limit is applied per velocity bin (or per wavelength), so the low flux wings of a peak, close to 0, will not be included in the analysis. This is especially important as the Balmer decrement can become arbitrarily high in noisy, dim regions of a spectrum.

We identified peaks in the intrinsic \ha{} spectrum using the function \texttt{scipy.find\_peaks}, which identifies local maxima by comparing to neighboring points, and recorded the $z$-velocity of each peak. The intrinsic spectrum was used to avoid identifying false peaks caused by scattering, which do not have any corresponding emitting particles along the sightline. Note that for observational data sets, this step is not possible; however, when peaks were found from the processed spectra, only 7\% of the identified peaks resulted from photons scattered into the sightline, so this is a minor effect.  We ensure that the peaks identified from the intrinsic \ha{} spectrum are still present and above the flux limit in the processed spectrum, after dust absorption and scattering is applied. Spectral components which disappear or are buried by the ISM after the effects of dust extinction are not included in the analysis. 
From these steps, we have a sample of peaks that would be visible if observed, but exclude the small number of peaks scattered into the sightline which would confuse the calculation of the $z$-position.  
For each peak present in the intrinsic and processed spectra, we find particles along the sightline at a $z$-velocity within 2.5\,km\,s$^{-1}$ of the peak, which would fall within 1 velocity bin at the resolution of the data cube. We create a histogram of particles' \ha{} luminosities against their $z$ positions, using 50 bins calculated from the $z$-position spread of the data so that more highly localized components can be better centered. From this histogram, the $z$ position which displays the highest luminosity is identified as the $z$ position of the group of particles which emitted that spectral component. 

In addition to the peak's velocity and $z$-position, we record their peak flux in the intrinsic and processed \ha{} and \hb{} spectra. The \hb{} flux is saved regardless of whether there was an identified peak in \hb{} at that velocity, to conserve peaks with a high Balmer decrement as part of the sample. Usually in observational studies, the Balmer decrement of a spectral component is calculated by summing the total flux of the line after fitting. As an approximation, we summed the line flux within 5\,km\,s$^{-1}$ of the center peak and used the resulting \ha{} and \hb{} values to calculate the ratio. The total dust mass along the sightline was also recorded, as well as the dust mass lying between the peak's $z$ position and the observer. Note that the recorded dust mass is the total amount of dust in a sightline based on the spatial resolution of 300\,pc, and therefore does not include the effects of strong variations in the dust surface density at small spatial scales. 

Some sightlines in the galaxy cover areas with disruptions such as supernovae blowing gas out the disk. These are evident in the spectra as regions for which there is no clear ISM component, or no component close to zero velocity or the central $z=0$ position. Since our method involves comparing outlying gas emission against the ISM component, we include only sightlines containing emission from the disk. The disk component is defined as a peak with a $z$ position within 0.2\,kpc of $z$=0 and a velocity between $-35$ and 35\,km\,s$^{-1}$. For a rough velocity cut of all peaks from within 0.2\,kpc of the disk at velocities within 50\,km\,s$^{-1}$ of 0, the standard deviation of the (relatively smooth) distribution was 17\,km\,s$^{-1}$, so the velocity cut used to identify the disk includes all peaks with velocities within 2$\sigma$ of 0. 
In addition to the ISM component, sightlines must contain at least one other component outside of the disk (based only on $z$-position, greater than 0.2\,kpc from the disk). 

The above steps are carried out for all sightlines at all timesteps, creating a sample of 510 standard sightlines containing an ISM component and an inflowing or outflowing component for comparison.

\section{Results} \label{sec:results}

We analyze these results by first examining a single line-of-sight (Section~\ref{subsec:casestudy}), and then looking at the distribution over all sightlines (Section~\ref{subsec:averages}).

\subsection{Case Study} \label{subsec:casestudy}

\begin{figure}
	\includegraphics[scale=0.85]{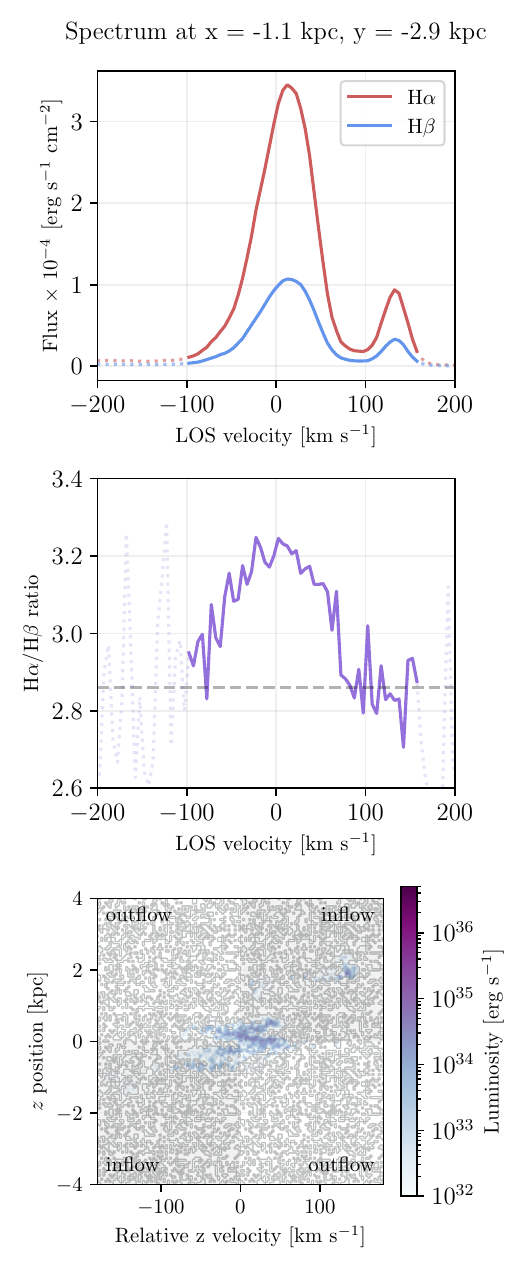}
    \caption{Top: \ha{} and \hb{} spectra of a single sightline $\sim3.1$\,kpc from the center of the galaxy in the earliest timestep. Dotted lines are regions below the flux limit which were left out of analysis, while solid lines are areas above the flux limit. Center: The Balmer decrement across the spectrum, with dotted lines again representing areas excluded from remaining analysis. The dashed gray line indicates the intrinsic Balmer decrement assuming Case B recombination and no dust. Bottom: A 2D histogram of the positions and velocities of particles in the sightline, weighted by luminosity. The displayed velocities are relative to the observer (not relative to the galaxy), so the sign of the velocity matches the spectra above.  See Section 4.1 for further details. }
    \label{fig:sightline}
\end{figure}

Figure \ref{fig:sightline} displays one example of the method we use to differentiate between inflow and outflow detected in \ha{} and \hb{}. This sightline lies 3.1\,kpc from the center of the galaxy in the earliest snapshot from the simulation ($t = 0.6$\,Gyr). It displays an ISM component at 12.5\,km\,s$^{-1}$ and a possible inflow or outflow component at 132.5\,km\,s$^{-1}$. The top panel displays the observed spectra of \ha{} and \hb{}. 

In the center panel, the \ha{}/\hb{} ratio at each velocity is displayed. The ISM component has a ratio of 3.31 (taking the average of the points within 5\,km\,s$^{-1}$ of the center of the peak), while the smaller component has a ratio of 2.87. The lower Balmer decrement for the high-velocity component indicates that this component should be in front of the disk. As the light passes through dust, the bluer wavelengths get absorbed more, increasing the \ha{}/\hb{} ratio. The ratio of the high-velocity component is very close to the intrinsic ratio of ionized gas at 10$^4$ K, 2.86, indicating that it has not traveled through very much dust. However, the ratio of the ISM component is higher, indicating that there is additional dust between this component and the other one, and that the ISM component must be behind the high-velocity component. The fact that the gas velocity is positive relative to the observer (meaning moving away), and the position is in front of the ISM according to the Balmer ratio, points to this component likely being inflow. 

The bottom panel allows us to identify inflows and outflows using particle information from the simulation. Particles which lie within the sightline are placed on a 2D histogram of their $z$ position and their $z$-velocity relative to the observer and weighted by their \ha{} luminosity. Since the $z$-velocity on the $x$-axis is relative to the observer, in order to match the velocities of the spectra above, the areas showing inflow have positive $z$-position and $z$-velocity (in front of the disk, and moving away from the observer), and areas showing outflow are negative position and velocity. These regions are shaded on the histogram. The ISM component lies at a $z$-position close to 0, and hosts a wide range of velocities centered around 0\,km\,s$^{-1}$. The wide range of velocities can be seen in the broadening of the line in the spectrum. The high-velocity component can be seen in the top right quadrant of the diagram, with a clump of bright gas at positive position (in front of the galaxy) flowing away from the observer and onto the galaxy (positive velocity). For this example, this method worked very well --- a high-velocity peak with a lower Balmer ratio correctly identified inflowing gas. However, we note that there are many sightlines that did not behave as expected, which are discussed further in the next section.

\subsection{Full galaxy} \label{subsec:averages}

\begin{figure*}
    \includegraphics[width=\textwidth]{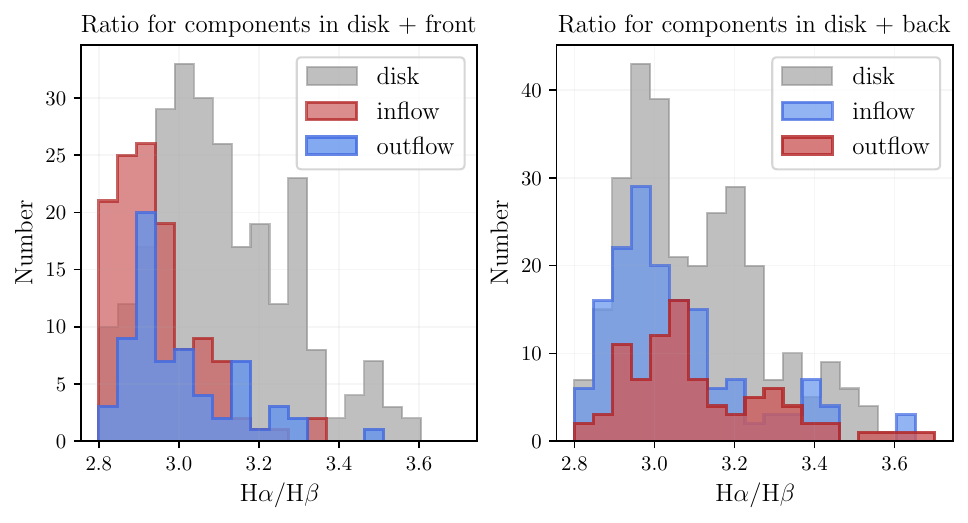}
    \caption{Left: The distribution of Balmer ratios between different components in the ISM and front of the galaxy. Red denotes inflow (red-shifted components in front of the galaxy) and blue denotes outflow. Both inflow and outflow peak at lower ratios than the disk. Right: A histogram showing the ratios of different components in the ISM and behind the galaxy. In this case, blue represents inflow (blue-shifted components behind the galaxy), and red represents outflow. Inflow peaks at a similar ratio to the disk, while outflow peaks at a ratio higher than the disk. }
    \label{fig:ratiohist}
\end{figure*}

Using the sample of sightlines with disk and inflow/outflow components above the flux limit, we now examine how the \ha{}/\hb{} ratio varies with $z$ position. Figure~\ref{fig:ratiohist} displays the distribution of \ha{}/\hb{} ratios for spectral components in different areas of the galaxy. The left panel shows the ratio distribution of sightlines with ISM and front components, while the right panel shows the ratio distribution of sightlines with ISM and back components. Note that the `disk' sample in each panel is not exactly the same, as most sightlines had only ISM and front or ISM and back components (though some had all three, and would be displayed in both panels). There are more sightlines with front components than back components, since components behind the disk in very dusty regions often lost the \ha{} peak due to extinction and so were not included in the sample. However, the distribution of ratios for the disk is similar for both histograms, spanning the entire 2.8-3.6 range and peaking at 3.03 and 2.97, respectively. 

As expected, components in front of the disk have overall lower ratios than components in the disk, as shown in the left panel. Inflowing and outflowing gas in the front of the galaxy displays ratios systematically lower than the disk component, with a peak shift of 0.1 and few components with ratios higher than 3.2. Though the peaks of inflowing and outflowing gas are the same, inflowing gas is distributed towards lower ratios than outflowing gas. 

We expect the components behind the disk would then exhibit higher ratios than the disk components due to dust absorption which happens between the disk and the components behind. In the right panel, outflowing components behind the disk peak at a ratio of 3.06 (0.1 higher than the disk component), though their range is similar. However, inflowing components have similar ratios to the disk components, peaking at 2.97 and displaying a similar range of ratios.

\begin{figure*}
    \includegraphics[width=\textwidth]{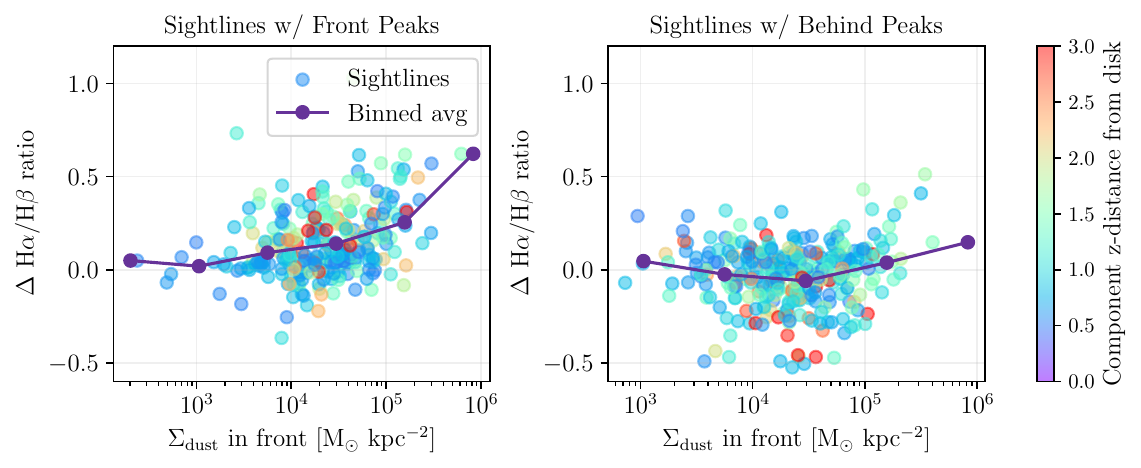}
    \caption{On the left, each point represents a sightline containing a peak from the disk and one from in front of the disk. The $x$ axis is the dust surface density including only the dust along the sightline between the two components, and the $y$-axis shows the \ha{}/\hb{} ratio of the front component subtracted from the \ha{}/\hb{} ratio of the disk component. On the right, the plot is the same but for sightlines containing peaks behind the disk, and the y axis is the ratio of the disk component subtracted from the ratio of the behind component. On each plot, the purple lines and dots display the average in different dust mass bins to make the trend (on the left) or lack of trend (on the right) visually clearer. }
    \label{fig:deltaplot}
\end{figure*}

In order to better understand how the Balmer decrement varies for components behind the disk, we plot the difference in the Balmer decrement of the two components against the difference in dust surface density between them, in Figure~\ref{fig:deltaplot}. Since we narrowed the sample to peaks containing an ISM component in addition to outlying gas, we can compare the ratio and the dust mass of the two components within the same sightline. Each point on the plot represents a single sightline, with the binned average displayed in purple. On the left, although there is a large amount of scatter, the peaks in the front display, on average, a larger change in ratio when there is more dust between them. On the other hand, for peaks behind the disk, in the right panel, the change in the ratio appears to be largely uncorrelated with the amount of dust lying between the disk and the outlying component, except at  high dust mass differences. Finally, we also color each peak by the outlying component's $z$-distance from the disk, and see that, somewhat surprisingly, there seems to be almost no correlation between the component's $z$-distance from the disk and the difference in dust mass along the sightline. 

\begin{figure*}
    \includegraphics[width=\textwidth]{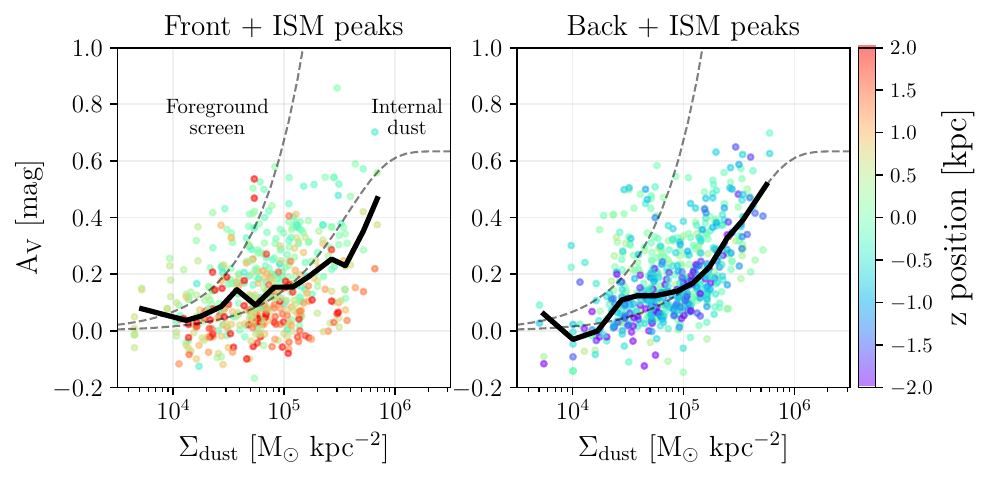}
    \caption{The dust extinction in the visual band A$_\mathrm{V}$ in magnitudes plotted against the surface density of dust $\Sigma_{\mathrm{dust}}$. Each point represents a peak from one of the spectra, colored by their $z$-position with respect to the disk. The solid black lines are running medians of the displayed points. The dashed gray lines represent theoretical curves of the 'foreground dust screen' and 'internal dust' models from \citet{calzetti1994} (see text for further explanation). The left side includes peaks from sightlines with front and ISM components, and the right side has peaks from sightlines with back and ISM components.}
    \label{fig:dustmodels}
\end{figure*}

In Figure \ref{fig:dustmodels}, we plot observable quantities --- A$_{\mathrm{V}}$, the extinction along the sightline in the visible band, and the dust mass surface density in each pixel $\Sigma_{\text{dust}}$. A$_{\mathrm{V}}$ was calculated from the Balmer decrement using the following equation:
\begin{equation}
    A_V = R_V \times \frac{1.086}{k(H\beta) - k(H\alpha)} \times \ln{ \frac{F(H\alpha)/F(H\beta)}{2.86} }
\end{equation}
with $F(\lambda)$ denoting the flux at the \ha{} and \hb{} wavelengths. We used a standard Milky Way extinction slope $R_V$ value of 3.1 and models of wavelength-dependent reddening $k(\lambda)$ from \cite{calzetti2000}. The dust surface density plotted in Figure \ref{fig:dustmodels} uses the total dust along the sightline, an observable quantity, rather than the dust between each component as in previous figures. A$_{\mathrm{V}}$ increases with dust along the sightline for peaks both in front of and behind the galaxy, as expected. The front peaks in this figure look very similar to Figure \ref{fig:deltaplot}, indicating that the total dust along the sightline mostly tracks with the dust in front of the sightline. 
On the other hand, while comparisons between the back components and the ISM in Figure \ref{fig:deltaplot} yielded no trend, A$_{\mathrm{V}}$ does actually increase with the dust as expected when counting all the dust in the sightline. 
This may indicate that there is large variation in the dust distribution happening below the chosen spatial $x$--$y$ scales of a single sightline, 0.3\,kpc. 

The points are also colored by their $z$-position along the sightline, with $z=0$ representing components in the disk. On the left side of the figure, redder points (indicating components which are further in front) tend to display slightly lower dust masses overall, with most of the high-dust mass components falling on the bluer end (denoting that they are in the disk). The right side of the figure shows that all different $z$ positions can have all different levels of dust, indicating that 1) dust is mixed in among the emitting components, and 2) there is lots of variation in the total amount of dust in each sightline. 
Different curves showing models of dust distribution and characteristics from \cite{calzetti1994} are overplotted in black, and will be discussed in Section \ref{sec:discussion}. 

While the emitting gas clumps in front behave in the expected way, those behind do not, especially inflowing gas. This could be caused by a clumpy dust distribution, which varies on a spatial scale smaller than the size of the sightline, and is more concentrated in the disk. Since the simulated galaxy displays energetic star formation, supernovae, and outflows in some locations, this may have stirred the dust into a clumpy medium distributed at many different scale heights across the galaxy. The simulation and radiative transfer itself takes into account the effects of small dust clumps, but co-adding spectra to a resolution of 0.3 kpc (detailed in Section \ref{sec:methods}) may have grouped dusty and non-dusty regions into one sightline.  

\begin{figure}
    \includegraphics[scale=0.85]{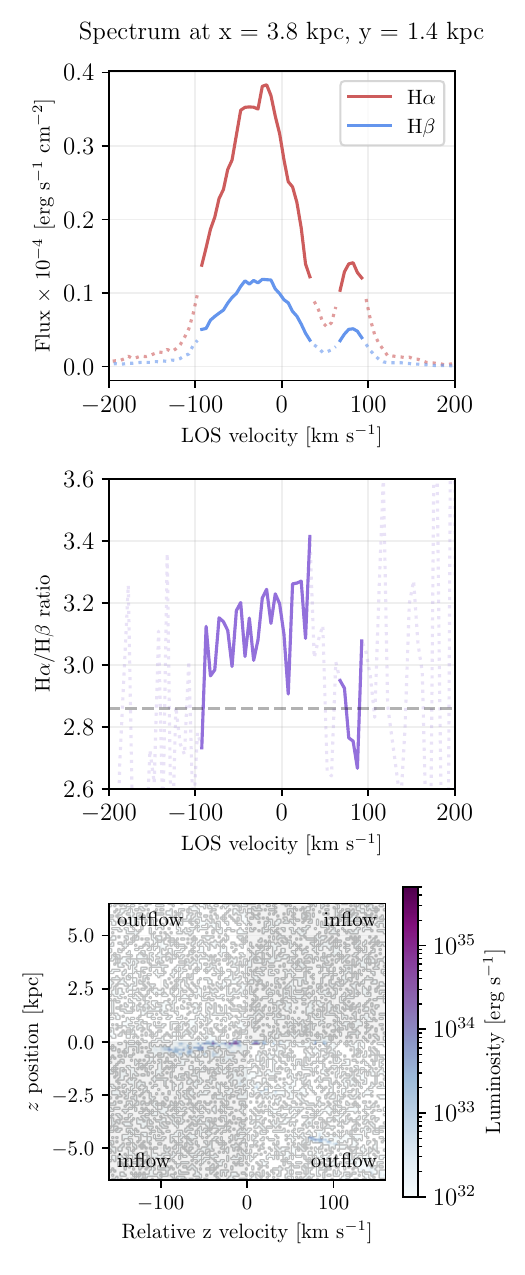}
    \caption{Top: \ha{} and \hb{} spectra of a single sightline $\sim$\,kpc from the center of the galaxy in the earliest timestep. Dotted lines are regions below the flux limit which were left out of analysis, while solid lines are areas above the flux limit. Center: The Balmer decrement across the spectrum, with dotted lines again representing areas excluded from remaining analysis. The dashed gray line indicates the intrinsic Balmer decrement assuming Case B recombination and no dust. Bottom: A 2D histogram of the positions and velocities of particles in the sightline, weighted by luminosity. The displayed velocities are relative to the observer (not relative to the galaxy), so the sign of the velocity matches the spectra above.}
    \label{fig:casestudy2}
\end{figure}

In order to explore how the dust distribution affects the Balmer ratio, we investigate a particular sightline which seems to show signs of dust clumpiness. The spectrum, displayed in Figure \ref{fig:casestudy2}, is emitted from a $0.3 \times 0.3$\,kpc bin positioned 4 \,kpc from the center of the galaxy, and displays a bright component peaking somewhere around $-17$\,km\,s$^{-1}$ and an additional fainter component at about 83\,km\,s$^{-1}$. In the second panel, the Balmer ratio of each component is displayed, with the lower-velocity disk component at around 3.2 and the higher-velocity component at 2.8. This would seem to indicate that the higher-velocity component is in front of the disk and (because it is at a positive velocity), inflowing onto the disk. However, the third panel shows that, in actuality, the positive-velocity component is outflow from behind the disk, despite the fact that it is at a lower Balmer decrement than the ISM component. 

Two possible effects may cause this: 1) there may be a clump of dust smaller than the size of the binned sightline, which lies in front of the ISM component and not in front of the back component, or 2) there may be blue light at 83\,km\,s$^{-1}$ scattered in from surrounding sightlines, lowering the ratio. Using the ``reddened'' spectral cube, which excludes scattering, we can ascertain that the ratio of the 83\,km\,s$^{-1}$ component remains lower.

\begin{figure}
    \includegraphics[width=\columnwidth]{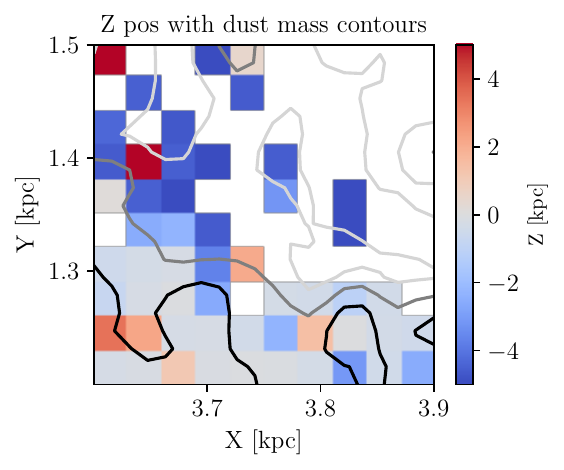}
    \caption{This is a zoom-in on a single sightline whose spectrum is displayed in Figure \ref{fig:casestudy2}. All pixels displayed fall within the same sightline. The colormap shows the z-position of the brightest emitting particles. The overlaid contours show the dust distribution, with the darkest contour denoting $1.2\times10^4$  $M_{\odot}$ kpc$^{-2}$, the middle contour $6\times10^3$  $M_{\odot}$ kpc$^{-2}$, and the lightest contour $2\times10^3$  $M_{\odot}$ kpc$^{-2}$.}
    \label{fig:sightlinedust}
\end{figure}

In Figure \ref{fig:sightlinedust}, we examine the positions of the emitting components and the positions of the dust in the sightline. All pixels shown in this figure fall within the sightline, and each pixel is colored by the average $z$ position of the bright particles in that $x$--$y$ bin. For visual clarity, we have plotted the z position of only the brightest 40\% of particles in the sightline. These particles alone re-create both peaks and 90\% of the total luminosity of the spectrum. The overlaid contours denote different dust surface densities (after smoothing), with the darkest contour indicating the most dust. The dust is clumped at very small scales, as regions of high dust and low dust fall within the same sightline. Bright ISM particles are present in the bottom left of the sightline, along with the heaviest concentration of dust. In the top left, there is bright emission from behind the disk (in blue) -- however, these photons travel through only a very thin dust layer. Because the dust is concentrated only in front of the ISM component, the ratio of the ISM component is higher despite the fact that it is in front.

While this specific sightline is an extreme example, this is likely the case for most sightlines which show a negative trend in ratio against $z$ position in the galaxy.  However, we note that the figures shown in Figure \ref{fig:sightlinedust}, which displays particle information at an $x$--$y$ scale finer than 0.3\,kpc, are approximations which treat particles as point sources positioned at their center. The simulation particles actually have varying sizes, many of which are larger than the $x$--$y$ scale shown in the figure (20\,pc), and these particles would therefore have their emission and dust surface density spread over multiple bins. This may smooth the emission and dust surface density slightly, but significant variation in the dust surface density still exists below the chosen observed resolution.

\section{Discussion} \label{sec:discussion}

In this section, we cover the observability of the difference in \ha{} and \hb{} in Section \ref{subsec:obs}, we discuss how realistic the extinction in the simulation is in Section \ref{subsec:ext}, and in Section \ref{subsec:simcaveats}, we consider how different properties of the simulation may affect our results.

\subsection{Observability} \label{subsec:obs}

\begin{table} \centering
 \caption{The results for the trends between the z-positions and Balmer ratios for red-shifted and blue-shifted components. In the middle column, the difference in the Balmer decrement between the velocity-shifted component and the ISM is given for red- and blue-shifted peaks originating from the front and back of the galaxy. In the right column, the percentage chance that a peak represents inflow is given for different observed Balmer decrements compared to the ISM.}
 \label{tab:ratios}
 \begin{tabular}{|c|c|c|c|c|}
  \hline
  & \multicolumn{2}{|c|}{$\Delta(\rm{H}\alpha/\rm{H}\beta{})$} & \multicolumn{2}{|c|}{\% Inflow Predicted} \\
  & \multicolumn{2}{|c|}{from z pos} & \multicolumn{2}{|c|}{from $\Delta(\rm{H}\alpha/\rm{H}\beta{})$} \\
  \hline
  & Front & Back & $|\Delta| > $ 0.1 & $|\Delta| > $ 0.3 \\
  \hline
  Redshifted & -0.20 & 0.01 & 82\% & 90\% \\
  Blueshifted & -0.11 & -0.05 & 75\% & -- \\
  \hline
 \end{tabular}
\end{table}

Using particle information from the simulation, we determined that emission from in front of the galaxy will have a lower Balmer decrement on average. This can be difficult to apply on the scale of an individual sightline. Using the same sample discussed above, if an offset component has a Balmer decrement lower than the ISM component, there is a 54\% chance that it originates from in front of the galaxy. An offset component with a Balmer decrement more than 0.3 below the ISM component has a 67\% chance of originating from the front of the galaxy. 

For emitting gas both in front of and behind the disk, outflows have higher ratios (as seen in Figure \ref{fig:ratiohist}), presumably because gas recently ejected from the galaxy will have more concentrated dust than inflowing gas from the outskirts. While dust-to-gas ratios for inflow and outflow are both quite low, averaging $\sim$0.6\%, outflows are slightly dustier when comparing only the brightest emitting particles ($\sim$0.7\% for inflows and $\sim$0.8\% for outflows). Due to this effect, the Balmer decrement can more easily be used to classify red-shifted components than blue-shifted components. The accuracy of the Balmer decrement in finding the component position and confirming inflow is summarized in Table \ref{tab:ratios}. In the right-most column, the percentage chance of the Balmer decrement correctly determining inflow is given for differences from the ISM component of 0.1 and 0.3. Note that these numbers are specifically for inflow -- for redshifted components this shows the percentage of peaks that are correctly identified as in front, and for the blueshifted components, the percentage correctly identified as behind the galaxy. The percentage of blue-shifted components from behind the galaxy identified using the Balmer decrement is not given as there were very few in that category. In the central column of the table, the average difference between the velocity-shifted and ISM components is given for front and back components. While front components have lower ratios for both red- and blue-shifted components, the trend is stronger for red-shifted components. For components from behind the galaxy, their ratio is on average similar to the ISM components. 

Observational studies have examined the difference in reddening between material around 0\,km\,s$^{-1}$ (assumed to be disk gas) and the broad wings of the spectrum (often assumed to be outflowing gas). Studies of different galaxies have yielded different results, with \cite{villarmartin2014} finding that, in quasars, the spectrum wings tended to be reddened with respect to the ISM, and \citet{mingozzi2019, Fluetsch2021, baron2024} finding that in Seyfert galaxies and ULIRGs, spectrum wings tend to be bluer than the ISM gas. This comparison is underexplored for Milky Way-like galaxies which are more similar to the simulated galaxy. While the range of $A_V$ from different components is greater in the observed galaxies than in the simulation, the differences of $\sim$0.3 in ratio ($\sim$0.2 in $A_V$) required to distinguish inflow from outflow should be observable by the same methods.

\subsection{Extinction} \label{subsec:ext}

In Figure \ref{fig:dustmodels}, we compare the extinction $A_V$ calculated from the Balmer decrement to the dust surface density of the simulation. The dashed gray lines plotted over the points represent the result of extinction and scattering from two different models of dust described in \cite{calzetti1994}: first, a foreground screen of non-scattering dust, and secondly, ``internal dust", where the emitting components are mixed in with the dust. Following \cite{calzetti1994} and \cite{kreckel2013}, we find that for a foreground screen, the extinction is described by:
\begin{equation}\label{eq:avscreen}
    A_{V,\text{screen}} = 0.67  \times \frac{\Sigma_{\text{dust}}}{10^5\,\text{M}_\odot/\text{kpc}^2} \, .
\end{equation}
Since this model excludes scattering, it represents one extreme of the library of dust models, in which the extinction rises exponentially with the dust surface density. On the other hand, for an internal scattering dust model, the extinction follows:
\begin{equation}
    A_{V,\text{internal}} = R_V  \times \frac{1.086}{k(H\beta) - k(H\alpha)} \times \ln {\frac{\gamma(H\alpha)}{\gamma(H\beta)}}
\end{equation}
\begin{equation}
    \gamma(\lambda) = \frac{1-e^{-\tau(\lambda)}}{\tau(\lambda)}
\end{equation}
\begin{equation}
    \tau(\lambda) = 0.921 k(\lambda)\sqrt{1 - w(\lambda)} \frac{A_{V,\text{screen}}}{R_V}
\end{equation}
\begin{equation}
    w(\lambda) = -0.48 \log(\lambda) \, ,
\end{equation}
with $\lambda$ being the wavelength of the line in Angstroms, and again using the standard Milky Way $R_V$ of 3.1 and the k($\lambda$) model in \cite{calzetti2000}. For this model, the scattering increases along with the extinction, creating an overall `grayer,' less wavelength-dependent effect on the emitted light which will cause the extinction to level off at a certain dust surface density. This represents the other possible extreme of the effect of dust on $A_V$. 

Observational studies have shown that in other galaxies, the distribution falls somewhere between the two models \citep{kreckel2013, tomicic2017}.  In our simulated galaxy, while the entire range of $A_v$ and dust matches this finding, the median of the points is centered on the internal dust model. The left panel of Figure \ref{fig:dustmodels} displays the front components and ISM components from those sightlines, and the right panel displays the back and components from those sightlines. The dust surface density is calculated from the dust along the entire sightline (what would be the observable quantity), so at each value of dust surface density there are at least two points at different $A_V$, denoting the ISM component and the front or back component. 

There is significant overlap between the two panels, which is expected as both include the ISM component. Both parts of the galaxy are more aligned with the internal dust model, supporting the idea that small clumps of dust dispersed among the emitting regions extinct and scatter the emitted photons. While largely overlapping in range, the back of the galaxy has a steeper upward slope, indicating higher total extinction in general. This may indicate that a large part of the attenuation occurs in the mid-plane of the galaxy, and provides support for the interpretation that emitted light from behind the galaxy only sometimes passes through small dust clumps in the disk on its path to the observer. The dust surface density is too low to see whether the ratio values will saturate as expected for the internal dust model.

While observations from \cite{kreckel2013} had higher overall dust surface densities, this may be due to the observed sample of galaxies which included some with prominent dust lanes, active galactic nuclei, or starbursts. Observations of different regions from M31 \citep{tomicic2017} displayed dust surface densities closer to those found in the simulation.

\subsection{Simulation Caveats} \label{subsec:simcaveats}

The simulation used in this work does not include an IGM or CGM surrounding the galaxy, and all inflows and outflows described are due to recycling gas. If IGM gas were accreting onto the galaxy, it might be expected that inflows would be less dust-enriched than outflows; this would widen the difference in ratio between red-shifted inflow and outflow, making the described methodology more useful for differentiating the two. However, it would narrow the difference between the average inflow/outflow ratios for blue-shifted peaks.

As the key results of this work rely on the predicted \ha{} and \hb{} emissivities, dust absorption plays a central role in determining the Balmer decrement. Although the simulations employed here include dust formation and destruction, the underlying dust physics remain highly simplified compared to complex nature of dust. Important processes — such as spatial variations in grain size distributions and compositions, dust–gas interactions via aerodynamic drag and Lorentz forces, and other mechanisms that govern grain growth and destruction — are not yet captured (e.g. \citet{draine2011, aoyama2018, mckinnon2018, mckinnon2019, li2019}). These omissions, in principle, can alter the shape of dust attenuation laws and thereby affect the predicted line emission. Future state-of-the-art galaxy simulations with more comprehensive dust modeling will be essential to quantify how different dust physics shape attenuation curves and impact emission-line predictions. Furthermore, while the adopted resolution ($\sim 10^3,M_\odot$) is sufficient to capture global star-formation activity, it remains challenging to resolve individual H II regions and the associated dust structures and clumpiness. The stochastic sampling of the initial mass function and the production of massive stars can be rare events in each spatial/velocity bins and therefore increase the scatter of the many relationships that we explored in this work. In future work, multiple simulations could be analyzed to disentangle the resolution and dust modeling effects, though this work represents a pilot study using a single simulation.

\section{Summary} \label{sec:summary}

In this paper, we have investigated whether the Balmer decrement (\ha{}/\hb{} ratio) of different components in an emission spectrum can be used to place those components along the observed sightline and differentiate inflow and outflow from a face-on galaxy. Dust in the disk of the galaxy should increase the Balmer ratio for spectral components behind the disk, while those in front of the galaxy should have a ratio closer to the intrinsic value. (See the schematic in Figure \ref{fig:cartoon} for an overview of the proposed method). We used an AREPO-RT simulation of a Milky Way-like galaxy and mock spectra created using COLT to test whether this method can work. Our findings are summarized below:

\begin{itemize}
  \item The \ha{}/\hb{} ratio of spectral components can be used to determine whether the gas is emitted from in front of the galaxy. Components emitted in front of the disk have a Balmer decrement averaging 0.14 lower than disk gas. However, because of the variation in dust mass across the galaxy, this method is best used to compare disk and outlying components in the same sightline. Spectral components which have a ratio more than 0.3 below the corresponding ISM component have a 67\% chance of being in front of the galaxy. (See Figure \ref{fig:ratiohist}). 
  \item Components from in front of the galaxy and inflowing components both tend to have lower \ha{}/\hb{} ratios. As a result, red-shifted components (which can be either front inflow or back outflow) can be located with respect to the disk more easily. Red-shifted components with ratios more than 0.3 below the corresponding ISM component have a 90\% chance of being in front of the galaxy, and any red-shifted component with a ratio below 3 has a 77\% chance of being in front of the galaxy. 
  \item The \ha{}/\hb{} ratio cannot be easily used to differentiate disk gas and gas from behind the galaxy, as they tend to have very similar distributions of ratios. This is because the dust in the simulation is distributed as small clumps (smaller than 0.3\,kpc, the size of a single sightline) which are more numerous in the disk. (See Figure \ref{fig:ratiohist} and Figure \ref{fig:sightlinedust}).
\end{itemize}

Future investigations into this method could use simulations with improved dust modeling and more dust-rich galaxies to amplify the effects of extinction.  

These results can be used to interpret spectra of face-on galaxies from IFU spectrographs such as \chas{} or MUSE. Large-scale inflow in galaxies has been rarely detected, and inflow may more commonly exist on smaller scales in specific regions of galaxies. The analysis method described and tested in this paper can help to reveal these smaller-scale inflows, analogous to our Milky Way's high- and intermediate-velocity clouds, that sustain star formation in galaxies.

\begin{acknowledgments}
We thank the anonymous reviewer for their helpful comments. MS was supported by NASA FINESST 80NSSC22K1603. HL is supported by the National Key R\&D Program of China No. 2023YFB3002502, the National Natural Science Foundation of China under No. 12373006 and 12533004, and the China Manned Space Program with grant No. CMS-CSST-2025-A10.
GLB acknowledges support from the NSF (AST-2108470, AST-2307419), NASA TCAN award 80NSSC21K1053, and the Simons Foundation through the Learning the Universe Collaboration.
We thank David Schiminovich and Nicole Melso for their input which improved the quality of the paper, as well as Ruby Sitaram for her encouragement. 
\end{acknowledgments}

\bibliography{example}{}
\bibliographystyle{aasjournalv7}



\end{document}